\documentclass[12pt]{iopart}

\usepackage{epsfig}
\usepackage{graphicx}
\usepackage[dvips]{color}
\usepackage{color}

\begin{document}

\title[The filamentation instability driven by warm electron beams]
{The filamentation instability driven by warm electron beams: 
Statistics and electric field generation}

\author{M E Dieckmann}

\address{Department of Science and Technology (ITN), Linkoping University, 
60174 Norrkoping, Sweden}
\ead{Mark.E.Dieckmann@itn.liu.se}

\begin{abstract}
The filamentation instability of counterpropagating symmetric beams of
electrons is examined with 1D and 2D particle-in-cell (PIC) simulations,
which are oriented orthogonally to the beam velocity vector. The beams are 
uniform, warm and their relative speed is mildly relativistic. The dynamics 
of the filaments is examined in 2D and it is confirmed that their 
characteristic size increases linearly in time. Currents orthogonal to the 
beam velocity vector are driven through the magnetic and electric fields 
in the simulation plane. The fields are tied to the filament boundaries 
and the scale size of the flow-aligned and the perpendicular currents are 
thus equal. It is confirmed that the electrostatic and the magnetic forces 
are equally important, when the filamentation instability saturates in 1D. 
Their balance is apparently the saturation mechanism of the filamentation 
instability for our initial conditions. The electric force is relatively 
weaker but not negligible in the 2D simulation, where the electron temperature 
is set higher to reduce the computational cost. The magnetic pressure gradient 
is the principal source of the electrostatic field, when and after the 
instability saturates in the 1D simulation and in the 2D simulation.
\end{abstract}


\maketitle

\section{Introduction}

The filamentation instability (FI) results in the growth and amplification
of magnetic fields in astrophysical plasmas and in laser-generated plasmas. 
Magnetic fields grow, for example, due to the redistribution of currents 
in an initially current-free system of two counterstreaming electron 
beams~\cite{Old1,Old2}. This is the simplest plasma configuration that
gives rise to the FI. The electronic FI grows faster than the competing 
two-stream and mixed mode instabilities, if the beams have a similar 
temperature and density and if their flow speed is at least mildly 
relativistic~\cite{New0,New1,New2}. A full classification of the competing 
modes in parameter space is given by Ref.~\cite{New3,Firpo}.
 
A wide range of previous numerical investigations of the FI exist, for
example the pioneering PIC simulation studies in 1D and in 2D 
systems~\cite{Old1,Old2}. Other important studies are the 1D Vlasov~\cite{Cal} 
and the 2D PIC simulations~\cite{Honda1} of counterstreaming electron beams, 
involving mobile ions. Relativistic beams and the impact of binary collisions 
on the FI have been investigated~\cite{Honda2}. The impact of a guiding 
magnetic field on the FI has been studied~\cite{Cary,StSchl,Stockem} as well 
as the combination of filamentation and Weibel instabilities~\cite{New4}. 
Equilibrium conditions of the beam-plasma system have been 
addressed~\cite{Hammer}. The statistical distribution of filament sizes has 
been the focus of 1D and 2D PIC simulation studies [16-19], while the 
astrophysical relevance of the FI driven by leptonic beams has been assessed 
in the Refs. [3, 20-26].

Here we revise the special case of the FI, which is driven by counterstreaming 
electron beams that are equally dense and moderately warm. This FI is 
electromagnetic during its linear growth phase \cite{New1,Firpo}. We model 
this FI with particle-in-cell (PIC) 
simulations. The growth and the saturation of the FI can be described as 
follows. Initially, the currents of the counterpropagating electron beams 
cancel each other. The noise inherent to PIC simulations results in 
fluctuating magnetic fields. A magnetic field fluctuation gives rise 
to a small separation of the electrons of both beams and, thus, to a 
net current. This net current amplifies, in turn, the magnetic field. 
The latter then grows exponentially until the FI saturates~\cite{Old1}. 

Simulations have evidenced the nonlinear growth of electrostatic fields 
by the FI and their importance has been pointed out~\cite{Cal,Honda2}. 
The comparison of the electric and magnetic field 
profiles~\cite{Stockem,Rowlands} for the case of symmetric beams, when 
no electrostatic field can grow in the linear phase of the FI, suggested, 
that the source mechanism is the magnetic pressure gradient force (MPGF). 
It induces through the acceleration of electrons a current, which 
drives the electrostatic fields. The fields oscillate around an equilibrium 
amplitude set by the MPGF~\cite{Brief} after the FI has saturated, unless 
positrons are present~\cite{Positron}. This current and the electrostatic 
fields are not affected by the introduction of a spatially uniform 
flow-aligned magnetic field due to its vanishing contribution to the
MPGF~\cite{Stockem}. A flow-aligned uniform magnetic field apparently only 
reduces the linear growth rate of the FI or it suppresses 
it~\cite{Cary,Stockem}. 

This paper is structured as follows. Section 2 outlines the simulation 
parameters. Section 3 revises a study~\cite{Brief} of the special case of 
symmetric cool electron beams. The link between the saturated electrostatic
field and the MPGF is illustrated and it is shown that the FI saturates 
due to an approximate cancelling of the magnetic and electrostatic forces. 
It is shown quantitatively, rather than 
qualitatively~\cite{Stockem}, for the first time that the saturated 
electric field in the simulation plane equals that expected from the MPGF 
in a 2D simulation. Section 5 is the summary.

\section{The PIC simulation method and the initial conditions}

The PIC method can, in principle, model all plasma processes in a 
collisionless plasma. It approximates the plasma phase space distribution 
by an ensemble of computational particles (CPs), each of which has a 
position $\bi{x}_{cp}$ and velocity $\bi{v}_{cp}$. Their charge to mass 
ratio equals here the $-e/m_e$ of an electron. The ions form in the 
simulations discussed here an immobile charge background, which cancels 
the net electron charge.

A PIC code solves the discretized Maxwell equations for the fields and the 
Lorentz equation for each CP and it interpolates the quantities defined
on the grid to the positions of the particles and vice versa~\cite{Dawson}. 
Our numerical scheme is outlined in Ref.~\cite{Eastwood}. The physical 
quantities are normalized as follows. The plasma frequency $\omega_p = 
{(2n_e e^2 / m_e \epsilon_0)}^{1/2}$ is obtained from the summed density 
of the two electron beams we model. Each beam has initially the spatially
uniform density $n_e$. The skin depth $\lambda_e = c / \omega_p$. The 
quantities in physical 
units denoted by the subscript $p$ are obtained from the normalized ones 
by substituting $\bi{E}_p = \omega_p c m_e \bi{E} / e$, $\bi{B}_p = \omega_p 
m_e \bi{B} / e$, $\bi{J}_p = 2en_e c \bi{J}$, $\rho_p = 2en_e \rho$, $\bi{x}_p 
=\lambda_e \bi{x}$, $t_p = t / \omega_p$, $\bi{v}_p = \bi{v}c$ 
and $\bi{p}_p = m_e c \bi{p}$. We also normalize $\Omega = \omega 
/ \omega_p$ and $\bi{k} = \bi{k}_p c / \omega_p$, where $\omega$
and $\bi{k}_p$ are given in physical units. The equations are 
\begin{eqnarray}
\nabla \times \bi{E} = -\frac{\partial \bi{B}}{\partial t},\,
\nabla \times \bi{B} = \frac{\partial \bi{E}}{\partial t}+\bi{J},
\, \nabla \cdot \bi{B} = 0, \, \nabla \cdot \bi{E} = \rho ,  \\
\frac{{\rm d}\bi{p}_{cp}}{{\rm d}t} = q_{cp} \left ( \bi{E}
[\bi{x}_{cp}] + \bi{v}_{cp} \times \bi{B}[\bi{x}_{cp}]\right ).
\end{eqnarray}
Initially $\bi{E}=0$ and $\bi{B}=0$. The electron beams have each a 
speed modulus $v_b = 0.3c$ and they move in opposite directions along 
$\bi{z}$. The thermal speed $v_e = {(k_B T_e / m_e)}^{0.5}$ is set to 
$v_b = 9v_e$ ($18v_e$) in the 2D (1D) simulation. The electron temperature
in the 2D simulation is higher to reduce the computational cost by the
increased Debye length, which determines both the grid cell size and the 
maximum time step that is possible. We resolve the x direction (1D) or the 
xy plane (2D), by which we isolate the FI with its $\bi{k} \perp 
\bi{z}$. The 2D simulation uses $1500 \times 1500$ grid cells to resolve 
the domain $L_x \times L_y = 90 \lambda_e \times 90 \lambda_e$. Each electron
beam is represented by 144 CPs per cell. The 1D simulation resolves $L_1 = 0.89 
\lambda_e$ by 500 grid cells and each electron beam by $1.21 \times 10^5$ CPs 
per cell like in Ref.~\cite{Brief}. The boundary conditions are periodic.

\section{Simulation results}

\begin{figure}
\centering
\includegraphics[width=0.49\columnwidth]{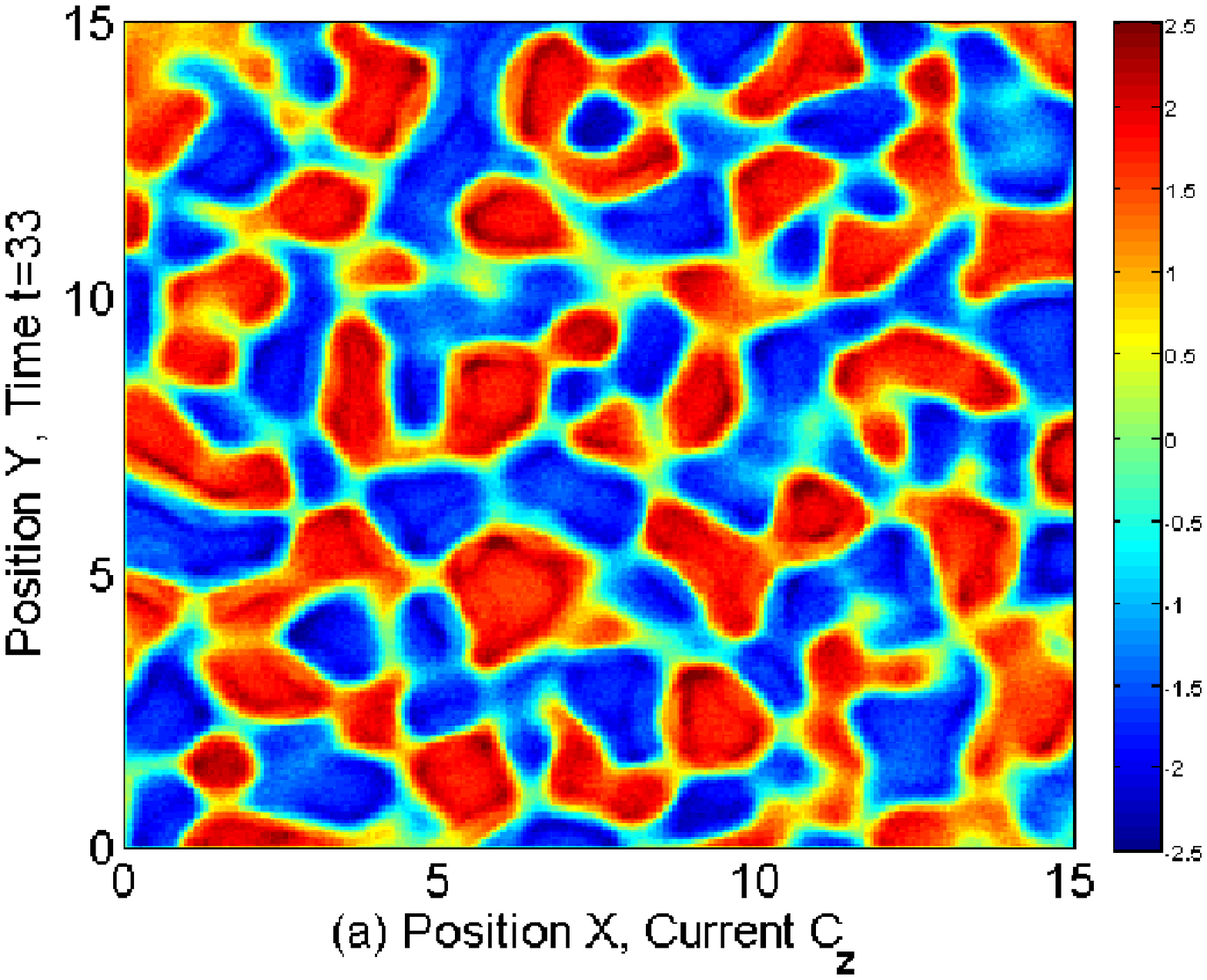}
\includegraphics[width=0.49\columnwidth]{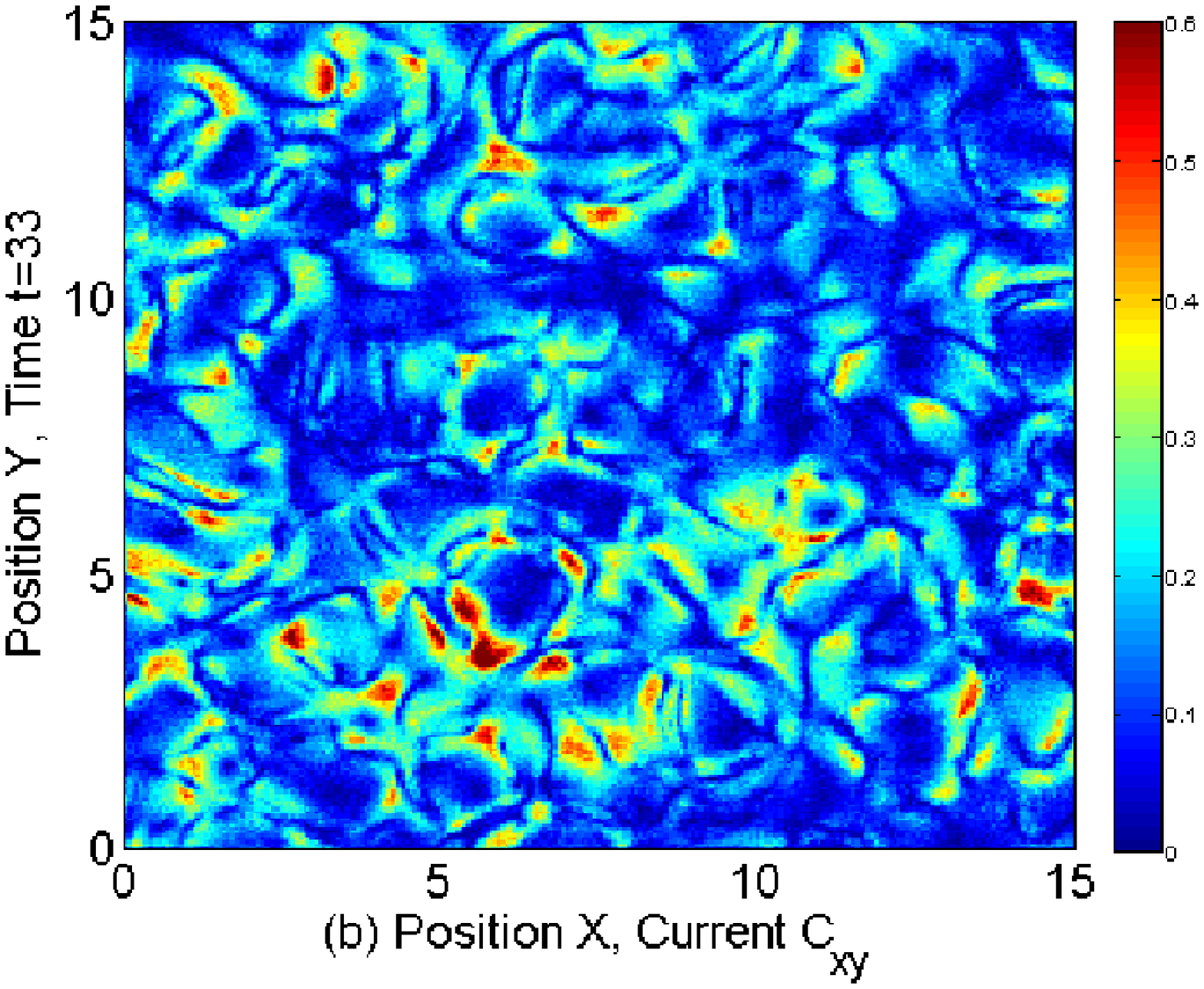}
\caption{(Colour online) The current in a section of the 2D box at $t=33$, 
normalized to the initial current of one beam: (a) displays the flow-aligned 
current $C_z$ and (b) the modulus of the current $C_{xy} = C_x + iC_y$ in the 
simulation plane. No obvious connection exists between the beam-aligned 
current and the perpendicular current.}\label{Plot1}
\end{figure}

The FI results in the separation of the currents $C_z(x,y)$ along $\bi{z}$
of both beams. We examine it and the current component in the (x,y)-plane, 
which we denote with $C_{xy}(x,y) = C_x(x,y) + i C_y(x,y)$. Both currents at 
the time $t=33$ in the 2D simulation are displayed in Fig.~\ref{Plot1}, when 
the FI has just saturated.
Both beams and their contributions to $C_z(x,y)$ have been separated into 
domains. The peak modulus is about twice the mean current of one beam. The 
structures in $|C_{xy}|$, which have a significant strength, show in some 
cases a correlation with those in $C_z$. The evolution of $C_z (x,y)$ is 
animated in time in the movie 1 and that of $|C_{xy}(x,y)|$ in the movie 2.
Movie 1 shows the initial growth of stationary filaments. These start
to merge and deform during the nonlinear phase of the FI. The filament
dynamics slows down in time, as the filament size increases relative to
the boundary speed. Movie 2 demonstrates, how structures in $|C_{xy}|$ 
come and go. They are damped and must thus be driven by $C_z$.

\begin{figure}
\centering
\includegraphics[width=0.49\columnwidth]{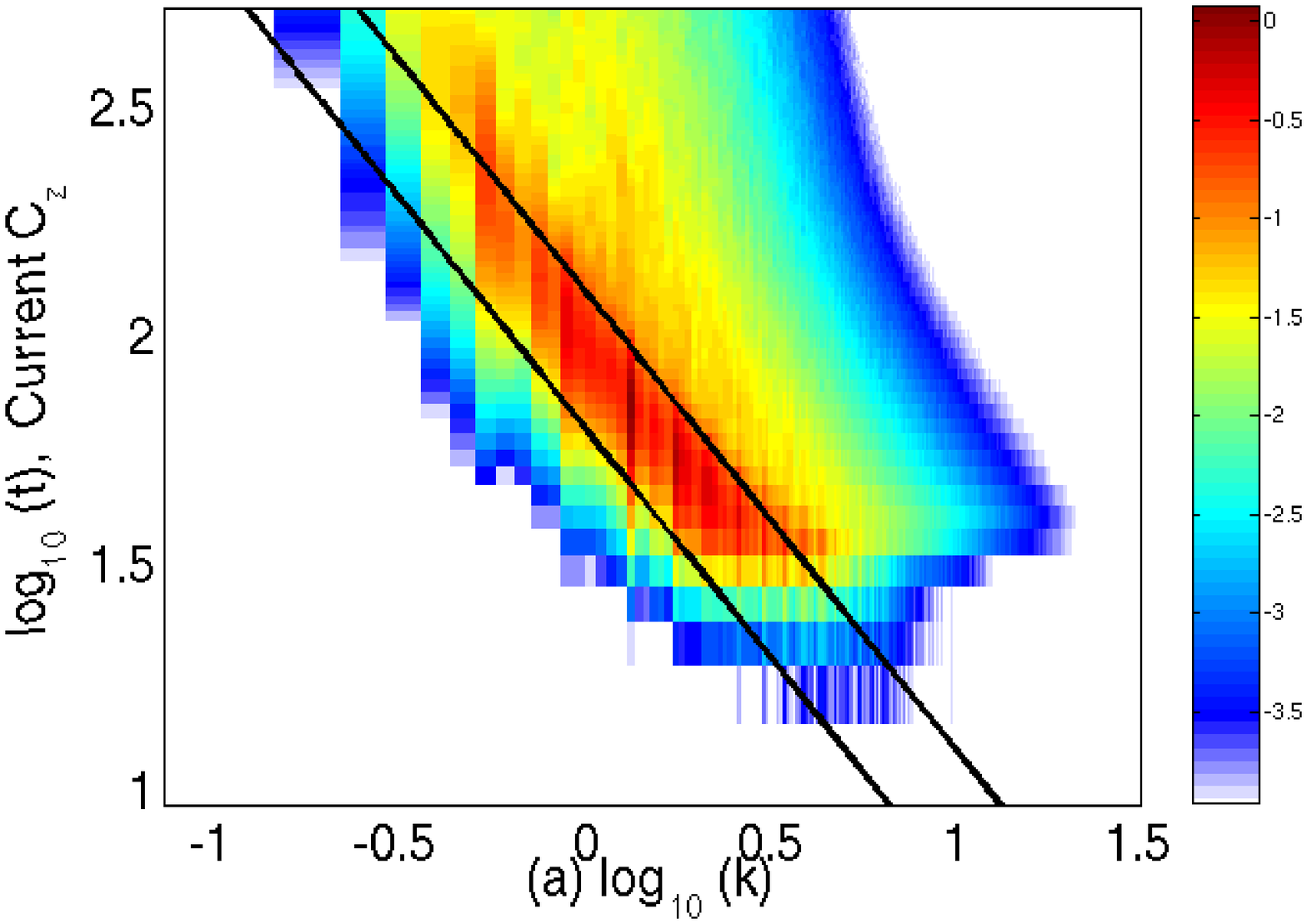}
\includegraphics[width=0.49\columnwidth]{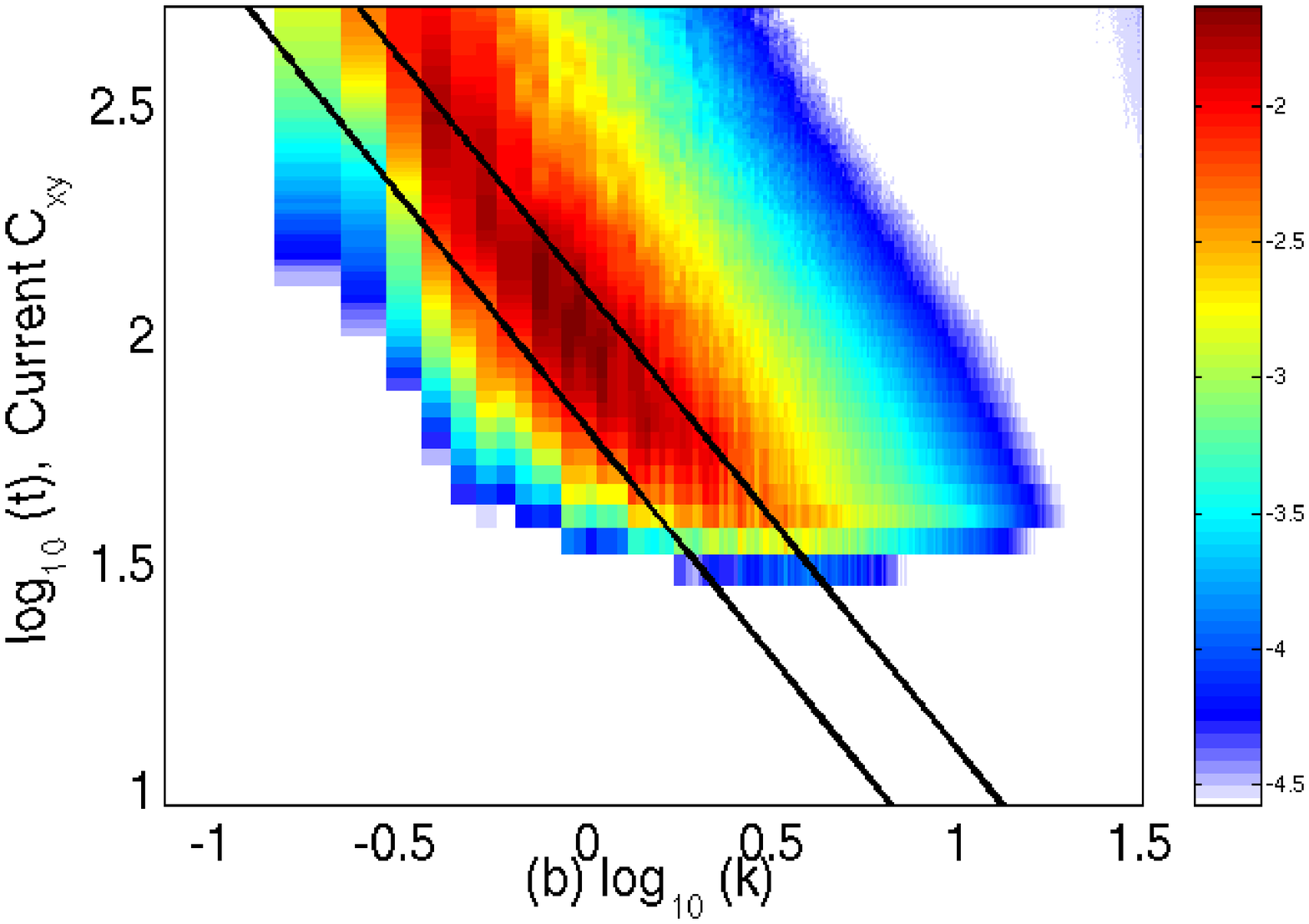}
\caption{(Colour online) The spatial power spectrum of the currents in 
the 2D box, integrated over the azimuth in $(k_x,k_y)$ as a function of
time: (a) corresponds to the $P_z(k,t)$ and (b) to $P_{xy}(k,t)$. 
Both power spectra are normalized to the maximum in (a) and the colour 
scale is 10-logarithmic. The same two curves $k \propto t^{-1}$ are 
overplotted.}
\label{Plot2}
\end{figure}

The $C_z(x,y,t)$ and the $C_{xy}(x,y,t)$ are Fourier transformed to 
$C_z (k_x,k_y,t)$ and $C_{xy}(k_x,k_y,t)$. The $P_z (k_x,k_y,t) 
= {|C_z(k_x,k_y,t)|}^2$ and $P_{xy} (k_x,k_y,t) = {|C_{xy}(k_x,k_y,t)|}^2$ 
are calculated. We transform $(k_x,k_y) \rightarrow (k \cos \alpha, k \sin 
\alpha)$ and integrate the power over $\alpha$ in the k-plane to give the
$P_z (k,t)$ and $P_{xy}(k,t)$ shown in Fig.~\ref{Plot2}.
The $P_z (k,t)$ is stronger than $P_{xy}(k,t)$, as expected. The k-interval, 
in which the power peaks in both current components, shifts in time like 
$k\propto t^{-1}$. The characteristic filament size $\propto k^{-1}$ thus 
increases linearly with the time. It is evident that the peak power of both 
spectra is located in the same k-interval at any fixed time, after the FI 
has saturated. The power $P_{xy}(k,t)$ maintains its slope at high $k$, 
while that of $P_z(k,t)$ is broadening in time. The shift of the boundary 
of $P_z(k,t)$ at low k to larger values of k at late times might be a finite 
box effect, as we notice the discreteness of k.

\begin{figure}
\centering
\includegraphics[width=0.49\columnwidth]{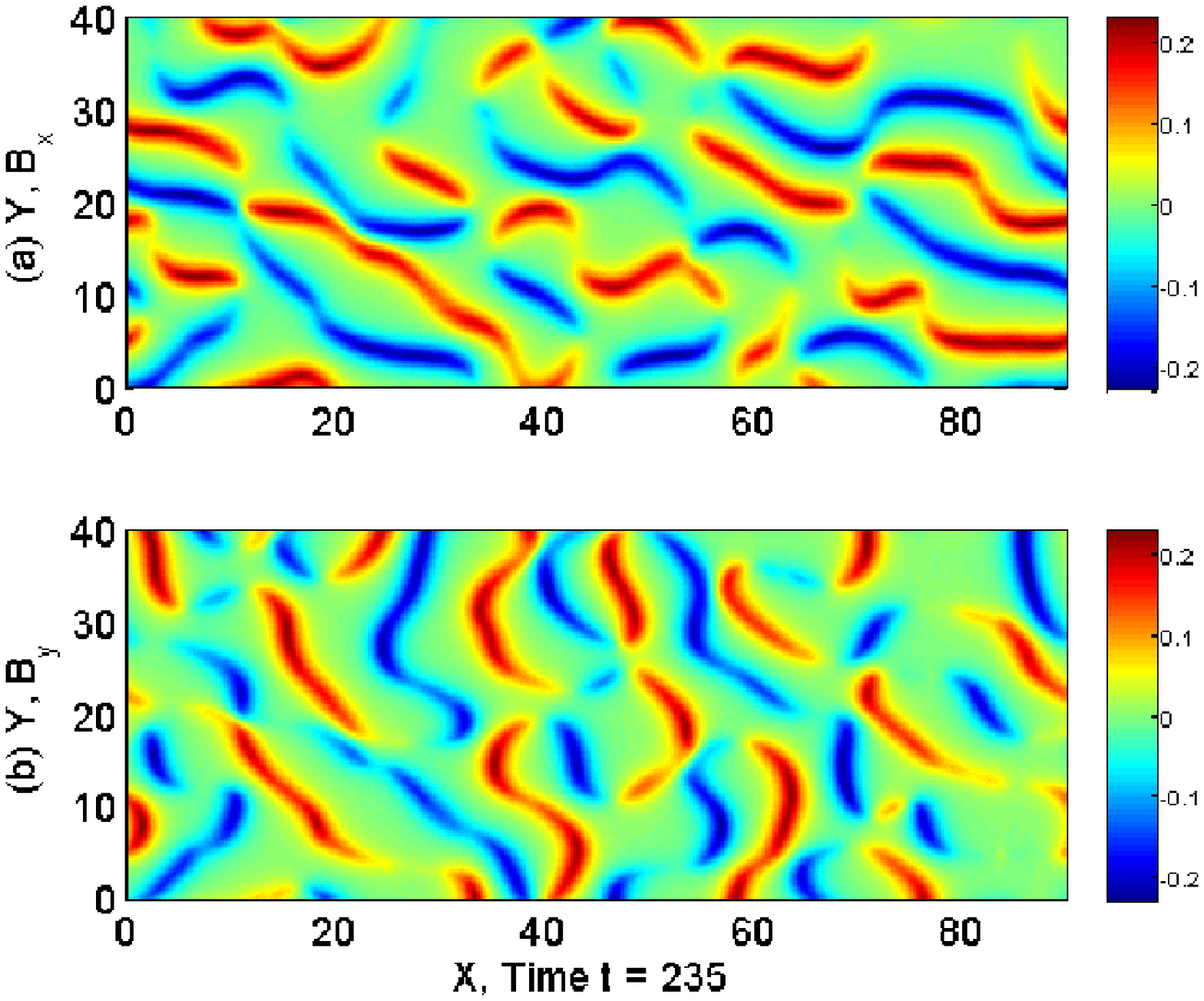}
\includegraphics[width=0.49\columnwidth]{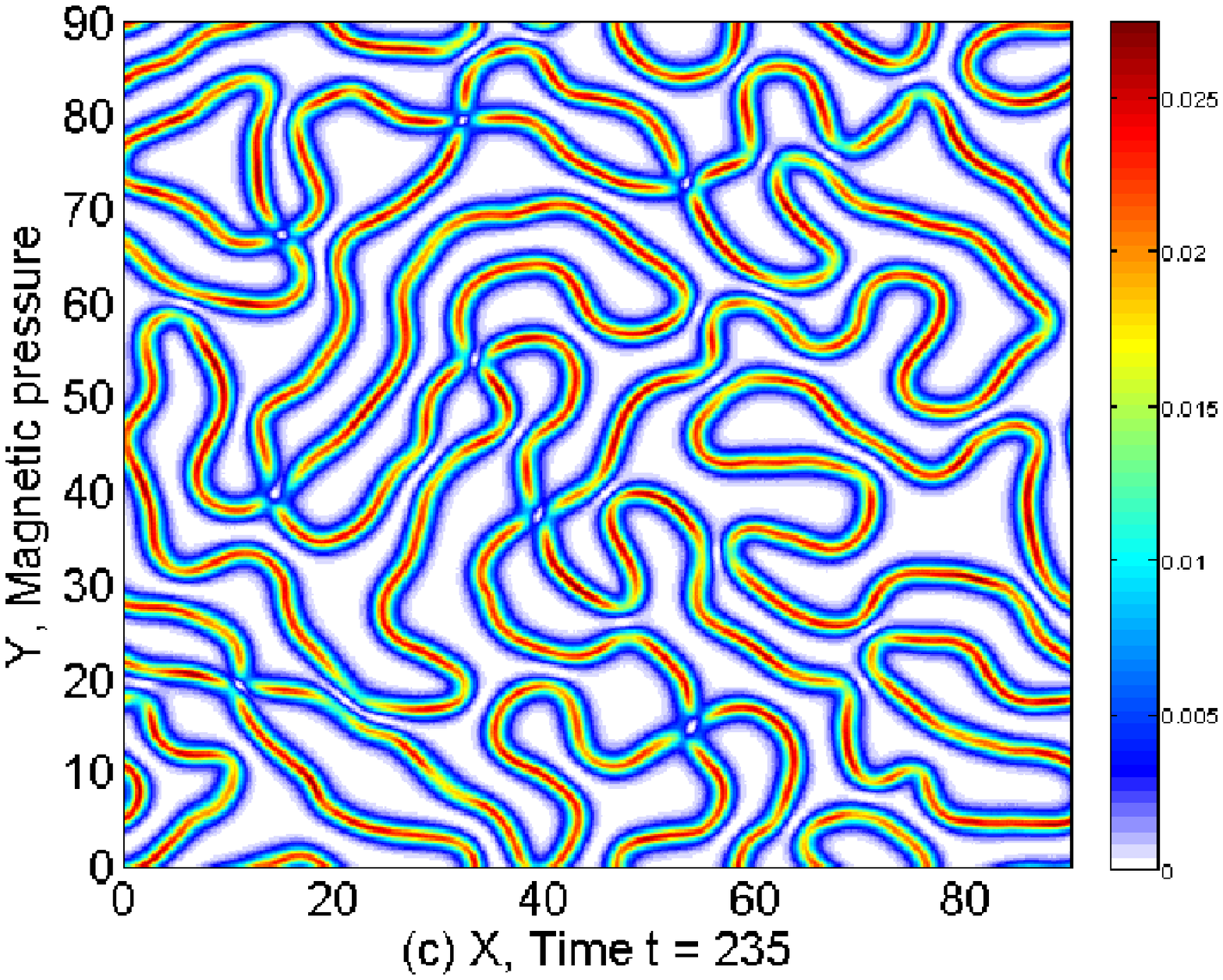}
\caption{(Colour online) The magnetic fields at $t=235$ in the 2D simulation: 
(a) displays the $B_x$ in a subsection of the simulation box and (b) 
the $B_y$ in the same interval. The magnetic pressure $P_B = 
(B_x^2 + B_y^2)/2$ in the full box is shown in (c).}\label{Plot3}
\end{figure}

Figure~\ref{Plot3} displays $B_x$, $B_y$ and the normalized magnetic pressure 
$P_B = (B_x^2 + B_y^2)/2$ at $t=235$.
The magnetic field vanishes within the filaments and it is strong at their
boundaries, as it is demonstrated by $P_B (x,y)$. The magnetic reconnection
points, e.g. at $x=12$ and $y=20$, and the locations ($x=20$, $y=20$), where 
reconnection is about to take place, demonstrate the merging of filaments. 
The $P_B (x,y)$ reveals strong gradients, which should result in a significant 
MPGF. 
A 1D PIC simulation provides more insight into the relevance of the MPGF. 
This can be exemplified with the help of the fluid momentum 
equation~\cite{Treumann}, which we will reduce for this purpose to one 
dimension $x$. Each species with the index $s$ is described by a fluid 
with the density $n_s$, the mean speed $\bi{v}_s$ and the pressure tensor 
$\mathbf{P}_s$ and the equation of motion in SI units is 
\begin{equation}
\partial_t (n_s \bi{v}_s) + \nabla \cdot (n_s \bi{v}_s \bi{v}_s) = 
-\frac{1}{m_e} \nabla \cdot \mathbf{P}_s + \frac{q_sn_s}{m_s}
\left ( \bi{E} + \bi{v}_s \times \bi{B} \right ).
\end{equation}
Ampere's law is used to rewrite $\bi{J}_s \times \bi{B}$ with $\bi{J}_s 
=q_sn_s\bi{v}_s$.  
We obtain the equation for the electron species $s=1$ and $s=2$, which
we model. 
\begin{equation}
\partial_t (n_s \bi{v}_s) + \nabla \cdot (n_s \bi{v}_s \bi{v}_s) = 
-\frac{\nabla}{m_e} \cdot \mathbf{P}_s - \frac{en_s}{m_e} \bi{E} - 
\frac{\nabla \cdot \bi{B}^2}{2 \mu_0 m_e} - \frac{\epsilon_0}{m_e} 
\bi{B} \times \partial_t \bi{E}\label{wampus} 
\end{equation}
The contribution by the gradient of the magnetic stress tensor $\mu_0^{-1} 
(\bi{B}\bi{B})$ is omitted. Its contribution is zero if we go to a 1D 
geometry, because the gradients along $y$ and $z$ vanish and because 
$B_x(x,t) = 0$. The thermal pressure tensor is diagonal for our initial 
conditions. The FI results in the 1D simulation in the initial growth of 
$B_y$ and $E_z$. The electrostatic $E_x$ grows nonlinearly. All other field 
components remain at noise levels. The component of Eq. \ref{wampus} along 
$x$ simplifies to
\begin{equation}
\partial_t (n_s v_{s,x}) +  \rmd_x (n_s v_{s,x}^2) = -\partial_x \frac{n_s k_B 
T_s}{m_e}-\frac{en_s}{m_e} E_x - \frac{B_y \rmd_x B_y}{m_e \mu_0}  
-\frac{\epsilon_0}{m_e} B_y \partial_t E_z.\label{wampus2}
\end{equation}
The electric and magnetic fields are computed self-consistently by the PIC 
simulation using the total charge and current density, which are obtained 
as $\rho = \rho_1 + \rho_2 + \rho_i$ and $\bi{J} = \bi{J}_1 + \bi{J}_2$ from 
the electron species 1 and 2 and from the background ions. The uniform and
constant charge density $\rho_i$ of the immobile ions cancels initially the 
electron charge and the ions do not provide a current. The $\rmd_x (n_s 
v_{s,x}^2)$ and 
the thermal pressure gradient can be neglected during the linear growth
phase, when only $B_y$ and $E_z$ grow. We retain $\partial_t (n_s v_{s,x}) 
\propto B_y d_x B_y$. The displacement current can probably be neglected, 
because of $B_y/E_z \approx 100$~\cite{Brief} and because $E_z$ grows
smoothly and aperiodically. A current grows along $x$ in response to the 
MPGF, which drives an electrostatic field through $\partial_t E_x = - J_x 
/ \epsilon_0$ or $\partial_t^2 E_x \propto -B_y d_x B_y$, because $\nabla 
\times \bi{B}=0$ in a 1D simulation.

The right hand side of Eq.~\ref{wampus2} is zero if $en_s E_x = -B_y \rmd_x 
B_y / \mu_0$ for the species $s$, if the thermal pressure gradient and the 
term with the displacement current are negligible. This may provide a 
condition for the saturation of the FI. The aperiodic growth of $E_x$ may 
be prescribed by $\partial_t^2 E_x \propto -B_y d_x B_y$ and the spatial 
amplitude of $E_x$ should in this case be proportional to that of the MPGF. 
An exact cancelling of the term 
$\propto E_x$ and the MPGF in Eq.~\ref{wampus2} is not possible, if $n_s$ 
varies as a function of $x$. The equations of both electron fluids are 
summed up to get a condition $(n_1+n_2) E_S = -2B_y \rmd_x B_y / \mu_0$ for 
the saturation electric field $E_S$ and we find that $n_1(x) + n_2(x) 
\approx 2n_e$ for this particular case study~\cite{Brief}. The normalization 
(section 2) to $E_S = -2B_y \rmd_x B_y$ facilitates the comparison of this 
feasible saturation condition with what we observe in the simulation. The 
factor 2 arises from the normalization to $2n_e$ in section 2. We define 
$E_B = E_S / 2$.

\begin{figure}
\centering
\includegraphics[width=0.48\columnwidth]{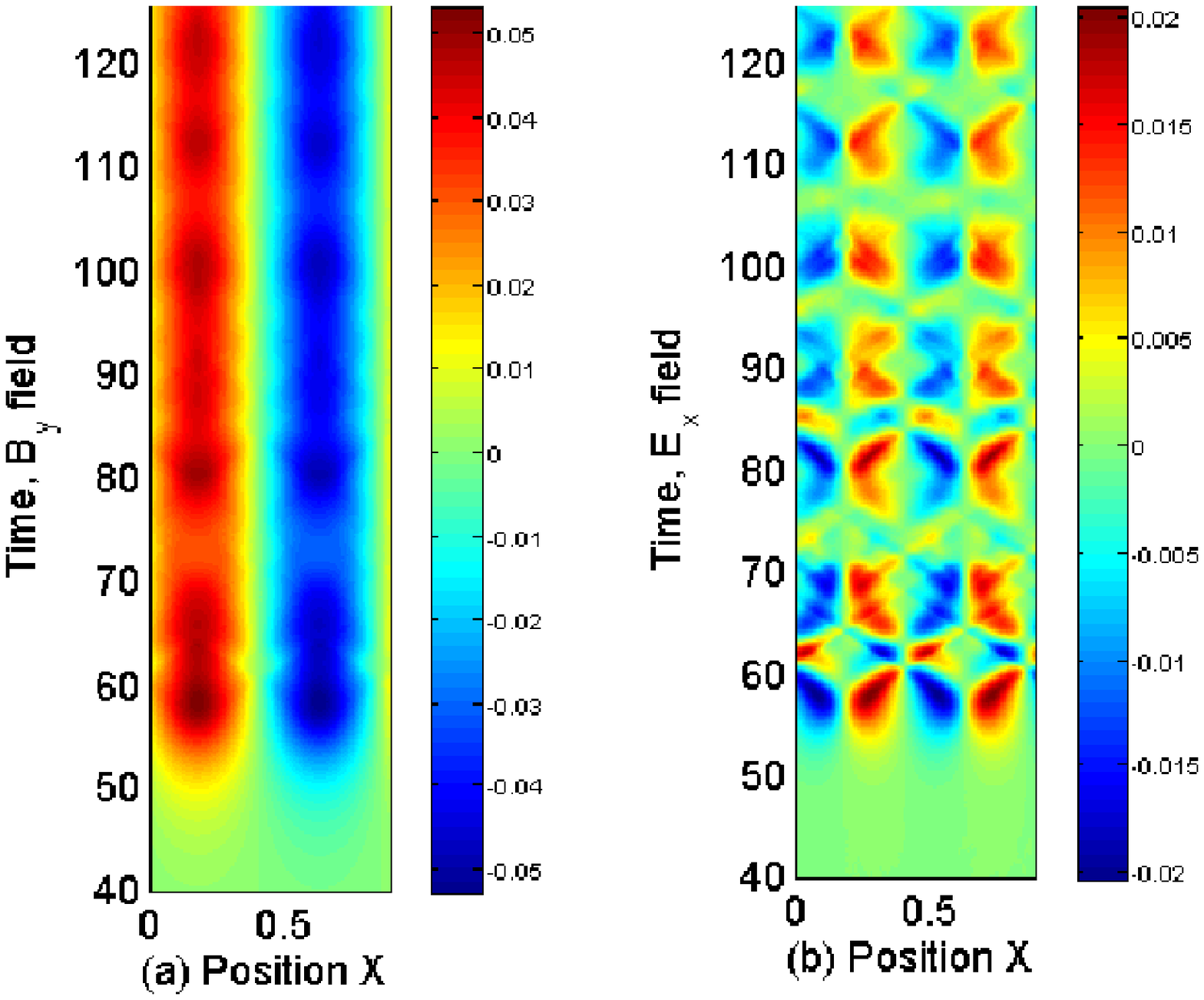}
\includegraphics[width=0.5\columnwidth]{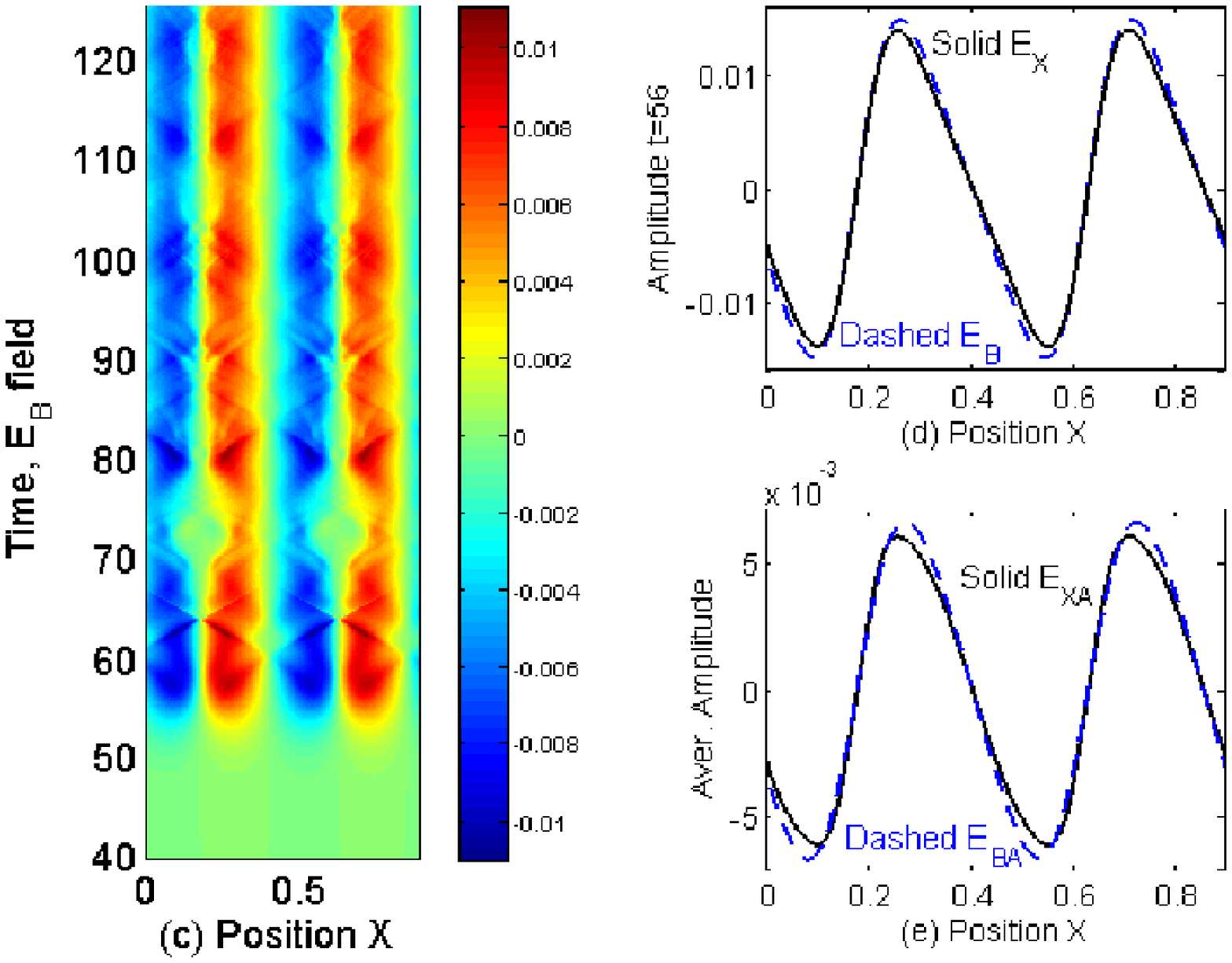}
\caption{(Colour online) The field data from the 1D simulation: (a) and (b) 
show $B_y(x,t)$ and $E_x(x,t)$, respectively. The $E_B (x,t)$ is displayed
in (c). The $E_x (x,t=56)$ (solid) and $2E_B (x,t=56)$ (dashed) are 
compared in (d), while (e) compares the time-averaged $E_{XA} (x)$ 
(solid)and $E_{BA} (x)$ (dashed).}\label{Plot5}
\end{figure}

Figure \ref{Plot5} compares the fields computed by the 1D PIC simulation.
Only one wave period of $B_y(x,t)$ is resolved by the box length $L_1$. 
The $B_y$ saturates at $t\approx 56$ and it remains approximately stationary 
thereafter. The saturation of $B_y$ is accompanied by the growth of $E_x$. 
The $E_x$ has twice the wavenumber of $B_y$ and it oscillates around a 
background electric field, which is stationary in space. The latter has the 
same wavenumber and amplitude as $E_B$. The $E_B$ and $E_x$ show correlations, 
e.g. at $t=63$. We demonstrate the quantitative match of $2E_B (x,t=56)$ 
and $E_x (x,t=56)$, when the latter reaches its peak amplitude. This accurate 
match confirms the dominance of the MPGF over the displacement current term
at this time, when the second and third term on the right hand side of 
Eq.~\ref{wampus2} practically cancel each other. A good agreement is also 
achieved between 
$E_{BA}(x) = {(t_2-t_1)}^{-1} \int E_B (x,t) \, dt$ and $E_{XA}(x) = {(t_2 
- t_1)}^{-1} \int E_x (x,t) \, dt$, which have both been integrated from 
$t_1 = 60$ to $t_2 = 130$. The $E_x(x,t)$ oscillates with the spatial 
amplitude $E_{BA}(x)$ around the stationary background field $E_{BA}(x)$. 
They add up to $E_x (x,t=56) = 2E_{BA}(x)$ and they cancel each other at 
the times $t_0$, when $E_x(x,t_0)=0$ in Fig. \ref{Plot5}(b). Ions will only 
weakly react to the high-frequency oscillation of $E_x$, but the stationary
$E_{BA}$ will accelerate them~\cite{Cal}. The ion current and its nonuniform
charge will eventually modify the fields and the force balance.

\begin{figure}
\centering
\includegraphics[width=0.47\columnwidth]{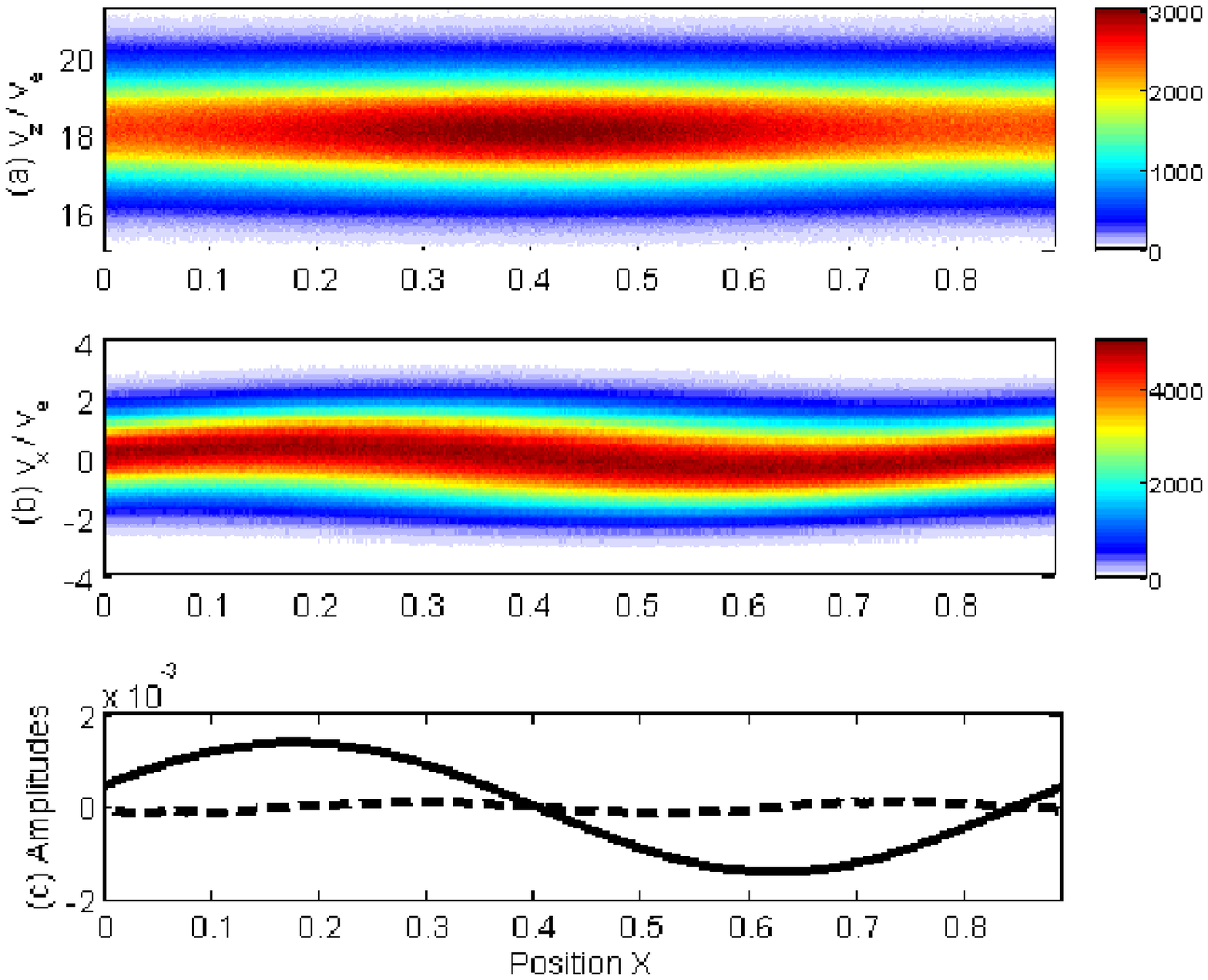}
\includegraphics[width=0.51\columnwidth]{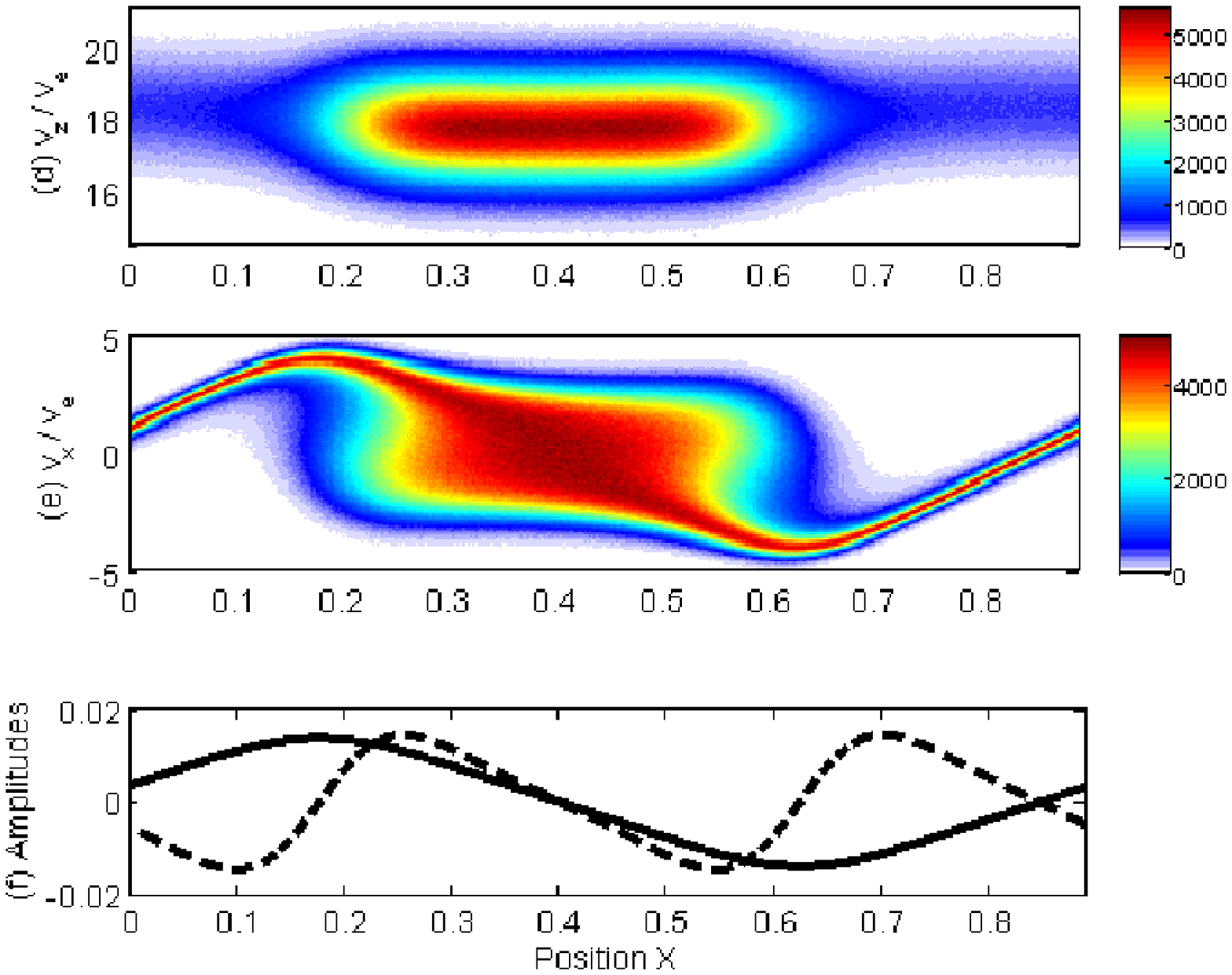}
\caption{(Colour online) Projections $f(x,v_z)$ and $f(x,v_x)$ of the 
electron phase space distributions and plots of the fields at $t=45$ 
(left column) and $t=56$ (right column). The colour scale of $f(x,v_z)$ 
in (a,d) and $f(x,v_x)$ (b,e) is linear and in units of CPs. The curves 
$v_b B_y$ (solid) and $E_x$ (dashed) are plotted in (c,e).}\label{Plot6}
\end{figure}

Figure~\ref{Plot6} compares the impact of $E_x$ and of $B_y$ on the 
saturation of the FI in the 1D simulation and on the phase space
distribution of the beam moving with $v=+v_b$ for the two times $t=45,
56$. 
The phase space distributions demonstrate that all electrons have a $v_z 
\approx v_b$ and a $|v_x| \ll v_b$. Movie 3 demonstrates that this is true 
throughout the simulation by animating the phase space densities $\log_{10} 
f(x,v_z,t)$ and $\log_{10} f(x,v_x,t)$. This weak heating may be the
reason, why the thermal pressure gradient has no obvious influence on
the saturation of the FI. Movie 3 reveals the rearrangement of the electrons 
of this beam into a filament and the vortex formation in $f(x,v_x)$. 

The normalized electric force on an electron is $-E_x$. The normalized 
magnetic deflection force working on an electron of the beam with 
$v_b > 0$ is $\approx -v_b B_y$. The magnetic force exceeds by far the 
electric one at $t=45$. Both forces are comparable at $t=56$, when the
FI saturates, which serves as a further illustration for the saturation
condition obtained from Eq.~\ref{wampus2}. The magnetic deflection force 
is responsible for the trapping 
of the filament electrons. The $E_x$ is repulsive at the centre of the 
filament and counteracts the magnetic trapping. It is attractive at longer 
distances, facilitating the filament overlap \cite{Brief}. The $E_x$ limits 
the peak density of the filament and, thus, the current it can carry 
\cite{Positron}. The combined action of both forces is to confine the 
electrons into a filament in $f(x,v_z)$ and a vortex in $f(x,v_x)$. The 
periodic oscillations of these electrons and their $J_x$ in the potential 
(Movie 3) give rise to the oscillatory $E_x$.

\begin{figure}
\centering
\includegraphics[width=0.49\columnwidth]{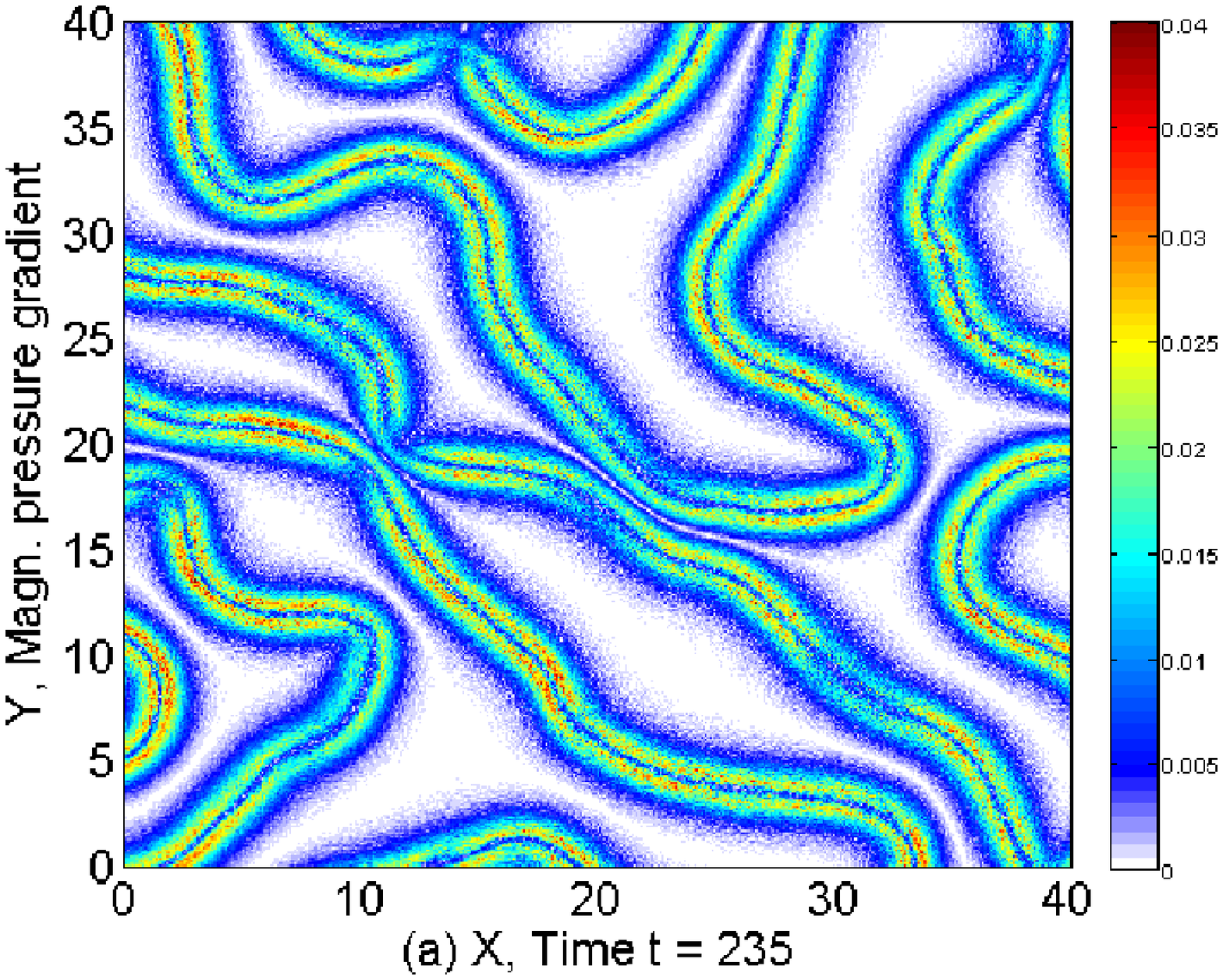}
\includegraphics[width=0.49\columnwidth]{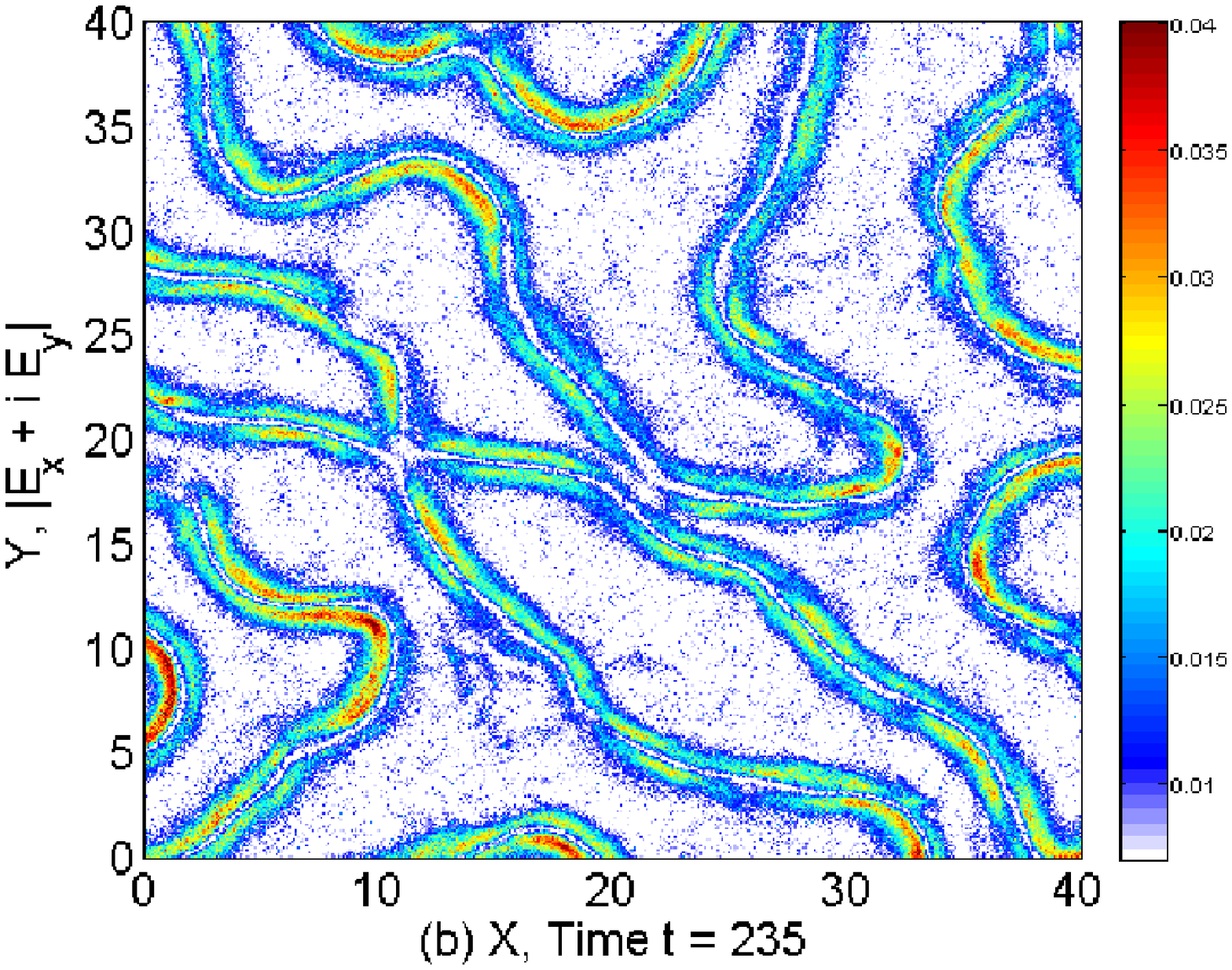}
\caption{(Colour online) The magnitude of $\bi{E}_B$ is shown for a 
subsection of the box 
at $t=235$ in (a). The electric field modulus $|E_x + iE_y|$ is displayed
in (b) for the same time and box interval. Both distributions have been 
computed by the 2D simulation.}\label{Plot4}
\end{figure}

Figure \ref{Plot4} compares the magnitude of $\bi{E}_B =-(d_x\bi{B}^2,
d_y\bi{B}^2)$ with the modulus of the complex electric field $E_x + iE_y$ 
in the 2D simulation at $t=235$. We select a late time, because then 
the filament dynamics is not so fast (movie 1) compared to the oscillation 
frequency of the electric field (Fig. \ref{Plot5}). The boundaries are also
quasi-planar, by which they become locally one-dimensional. This should
reduce the importance of the magnetic tension force relative to the MPGF.
The $|\bi{E}_B|$ and $|E_x + iE_y|$ are spatially correlated, they have the 
same magnitude and they reveal a split in the centre of the band. This split 
occurs in the 1D simulation, when $d_x B_y = 0$. It is thus likely, that
the electric field in the simulation plane is driven practically
exclusively by the MPGF. Both fields are out of phase at some places, e.g. 
at $(x,y)=(10,10)$, where $|\bi{E}_B|$ does not have a maximum. This may be 
explained with the interplay of $\bi{J}$ with $\bi{E}$, causing oscillations 
of the electric field. The peak modulus of $|\bi{E}_B| \approx 0.04$, which 
is about half the peak value of the magnetic deflection forces $v_b |B_x|$ 
or $v_b |B_y|$ in Fig.~\ref{Plot3}.

\section{Discussion}

We have investigated the FI with 1D and 2D PIC simulations, where the
beam velocity vector is orthogonal to the simulation direction or plane.
The uniform, cool and symmetric electron beams have a mildly relativistic
relative speed. The 2D simulation has provided insight into the interplay
of the filaments, e.g. their merging, during the nonlinear phase of the FI.
The filaments formed by the beam-aligned current couple to the perpendicular
currents through the electromagnetic fields. The fields are confined to the 
filament boundaries and the characteristic size of the structures in the 
flow-aligned and the perpendicular current are thus equal. This characteristic 
size increases linearly with the time. The power spectrum of the damped
perpendicular currents showed for all times after the FI has saturated a 
constant slope at large wavenumbers, which follows approximately a power 
law, albeit over a narrow range of wavenumbers \cite{Dieckmann}. 

The 1D PIC simulation using the same setup as in Ref. \cite{Brief} 
revealed a feasible source mechanism for the electrostatic fields, which 
develop during the nonlinear phase of the filamentation instability. 
The MPGF accelerates the electrons and a current develops, which couples 
through Amperes law into the electrostatic field. This mechanism is
suppressed by selecting beams of equally dense electrons and positrons,
because here the currents of both species cancel \cite{Positron}. The
growth of the electrostatic fields is not affected by a spatially uniform 
and flow-aligned magnetic field, because its MPGF contribution 
vanishes~\cite{Stockem}. It can reduce the growth rate of the FI though.

The current and the electric field oscillate in the 1D simulation after 
the FI has saturated. The spatial profile of the time-averaged electrostatic
field amplitude is that expected from the MPGF. The electric field amplitude 
oscillates between the initial value $E_x(x,t=0) = 0$ and twice its  mean
amplitude. The nonlinear terms in a 1D fluid equation due to the electrostatic
field and the MPGF cancel approximately, when the electrostatic field peaks.
This may suggest that, for the special case of symmetric and nonrelativistic 
electron beams considered here, the FI saturates due to the balancing
of the electrostatic and magnetic forces. We demonstrated for the first time 
that the MPGF is also responsible for the electric field in the simulation
plane of the 2D PIC simulation. The electric field amplitude in the 2D
simulation is weaker than that in the 1D simulation, but it is not negligible.
This might be a consequence of the increased electron temperature that 
reduced the computational cost of the 2D simulation. The magnetic energy
density we obtain is about 5\% (10\%) of the total energy in the 1D (2D)
simulation. The magnetic field amplitude for a plasma frequency $\approx$ 
10 kHz, which is representative for the solar wind, would be about 20 nT.

Our findings will be relevant for (almost) symmetric electron beams.
Selecting asymmetric beams facilitates plasma equilibria~\cite{Hammer}, 
rather than a time-dependent evolution. The FI will then also result in 
the generation of electrostatic fields during its linear growth 
phase~\cite{New1} and the mixed mode instability will outrun the FI. 
Highly relativistic beams will probably break the simple relation between 
the MPGF, the current and the electrostatic field. The thermal pressure 
gradient and the magnetic tension may not be negligible for other initial 
conditions and during the initial nonlinear stage, when the filament boundary 
curvature is higher. The competition of the FI with the mixed mode instability 
and the electrostatic two-stream instability, which saturates by forming 
phase space holes, as well as the flux tube bending, will introduce 
nonlinear effects that can only be addressed with large-scale 3D PIC and
Vlasov simulations.

{\bf Acknowledgements} The author would like to thank Vetenskapsr\aa det, 
the Deutsche Forschungsgemeinschaft (FOR1048) and the HPC2N for their support.


\begin{thebibliography}{}

\bibitem{Old1} Davidson R C, Hammer D A, Haber I and Wagner C E
1972 {\it Phys. Fluids} {\bf 15} 317

\bibitem{Old2} Lee R and Lampe M 1973 {\it Phys. Rev. Lett.} {\bf 31} 
1390

\bibitem{New0} Schlickeiser R and Shukla P K 2003 {\it Astrophys. J.}
{\bf 599} L57 

\bibitem{New1} Tzoufras M, Ren C, Tsung F S, Tonge J W, Mori W B, 
Fiore M, Fonseca R A and Silva L O 2006 {\it Phys. Rev. Lett.} {\bf 96}
105002

\bibitem{New2} Bret A, Gremillet L and Bellido J C 2007 {\it Phys. Plasmas}
{\bf 14} 032103

\bibitem{New3} Bret A, Gremillet L, Benisti D and Lefebvre E 2008
{\it Phys. Rev. Lett.} {\bf 100} 205008 

\bibitem{Firpo} Bret A, Firpo M C and Deutsch C 2004 {\it Phys. Rev. E}
{\bf 70} 046401

\bibitem{Cal} Califano F, Cecchi T and Chiuderi C 2002 {\it Phys. Plasmas} 
{\bf 9} 451 

\bibitem{Honda1} Honda M, Meyer-ter-Vehn J and Pukhov A 2000
{\it Phys. Rev. Lett.} {\bf 85} 2128 

\bibitem{Honda2} Honda M, Meyer-ter-Vehn J and Pukhov A 2000
{\it Phys. Plasmas} {\bf 7} 1302 

\bibitem{Cary} Cary J R, Thode L E, Lemons D S, Jones M E and Mostrom M A 1981
{\it Phys. Fluids} {\bf 24} 1818

\bibitem{StSchl} Stockem A, Lerche I and Schlickeiser R 2007
{\it Astrophys. J.} {\bf 659} 419

\bibitem{Stockem} Stockem A, Dieckmann M E and Schlickeiser R 2008
{\it Plasma Phys. Controll. Fusion} {\bf 50} 025002

\bibitem{New4} Lazar M, Schlickeiser R and Shukla P K 2006 {\it Phys. Plasmas}
{\bf 13} 102107 

\bibitem{Hammer} Hammer D A and Rostocker N 1970 {\it Phys. Fluids} {\bf 13}
1831

\bibitem{Rowlands} Rowlands G, Dieckmann M E and Shukla P K 2007
{\it New J. Phys.} {\bf 9} 247

\bibitem{Medvedev} Medvedev MV, Fiore M, Fonseca RA, Silva LO and 
Mori WB 2005 {\it Astrophys. J.} {\bf 618} L75

\bibitem{Dieckmann} Dieckmann M E, Lerche I, Shukla P K and Drury L O C
2007 {\it New J. Phys.} {\bf 9} 10

\bibitem{SilvaAIP} Silva L O 2006 {\it AIP Conf. Proc.} {\bf 856} 109

\bibitem{Petri} Petri J and Kirk J G 2007 {\it Plasma Phys. Controll. Fusion}
{\bf 49} 297

\bibitem{Sakai} Sakai J I, Schlickeiser R and Shukla P K 2004 {\it Phys.
Lett. A} {\bf 330} 384

\bibitem{Kazimura} Kazimura Y, Sakai J I, Neubert T and Bulanov S V
1998 {\it Astrophys. J.} {\bf 498} L183

\bibitem{Fonseca} Fonseca R A, Silva L O, Tonge J W, Mori W B and 
Dawson J M 2003 {\it Phys. Plasmas} {\bf 10} 1979 

\bibitem{SilvaAPJ} Silva L O, Fonseca R A, Tonge J W, Dawson J M, 
Mori W B and Medvedev M V 2003 {\it Astrophys. J.} {\bf 596} L121

\bibitem{Jaroschek} Jaroschek C H, Lesch H and Treumann R A 2004
{\it Astrophys. J.} {\bf 616} 1065 

\bibitem{Positron} Dieckmann M E, Shukla P K and Stenflo L 2009
{\it Plasma Phys. Controll. Fusion} {\bf 6} 065015

\bibitem{Brief} Dieckmann M E, Kourakis I, Borghesi M and Rowlands G
2009 {\it Phys. Plasmas} {\bf 16} 074502

\bibitem{Dawson} Dawson J M 1983 {\it Rev. Mod. Phys.} {\bf 55} 403

\bibitem{Eastwood} Eastwood J W 1991 {\it Comput. Phys. Commun.} {\bf 64} 252

\bibitem{Treumann} Baumjohann W and Treumann R A 1996 {\it Basic Space Plasma
Physics} (London: Imperial College Press) p 138

\end{thebibliography}
\end{document}